# Creating molecular structures from bilayer graphene and other materials

Laith Algharagholy, Steven W.D. Bailey, Thomas Pope and Colin J. Lambert*

Department of Physics, Lancaster University, Lancaster LA1 4YB, UK

Address correspondence to c.lambert@lancaster.ac.uk

**We demonstrate a new technique for creating unique forms of pure sp$^2$-bonded carbon and unprecedented hetero-molecules. These new structures, which we refer to as** *sculpturenes,* **are formed by sculpting selected shapes from bilayer graphene, hetero-bilayers or multi-layered materials and allowing the shapes to spontaneously reconstruct. The simplest sculpturene is topologically equivalent to a torus, with dimensions comparable to those of fullerenes. The topology of these new molecular structures is stable against atomic scale defects.**

Key words: nanostructures, self-assembly, directed assembly, graphene, carbon nanotubes

Although a variety of techniques have been developed for synthesising fullerenes[1-3], carbon nanotubes (CNTs)[4-8] and more exotic structures such as CNT Y-junctions, T junctions, toruses, bamboos, buds and horns[4], their synthesis is invariably uncontrolled. For example, CNTs grown using arc discharge, laser ablation or chemical vapour deposition typically possess a mixture of chiralities and the more exotic structures usually appear only by chance. On the other hand, the synthesis of graphene nanoribbons (GNRs) is more deterministic and can be achieved using lithographic[9-11], chemical[12-15] and sonochemical[16,17] techniques, unzipping CNTs[18-24] and assembling GNRs from chemical precursors[25]. For example, STM lithography[26] can be used to cut GNRs with widths as small as 2.5nm, with a specified chirality, a specified location and with their ends contacted to (graphene) electrodes. However, unlike fullerenes and closed CNTs, GNRs are not pure forms of sp$^2$-bonded carbon, because sp$^2$ bonding is disrupted by their edges, which are difficult to control at an atomic scale[24]. This disruption is undesirable, because as a consequence, for a given electron energy gap, the mobility of GNRs is typically significantly lower than that of CNTs[24,27-32]. Efforts to control this disruption have led to rapidly-increasing interest in the physics and chemistry of graphene edges and in particular, edge coalescence (EC) in multilayer structures, through which open edges reconstruct to form closed edges with a higher degree of sp$^2$ bonding. Open edges and ECs in bilayer graphene have been imaged by a number of groups[44,47]. After thermal annealing or Joule heating, open edges are rarely found and the more frequent ECs are observed to run mainly along achiral (zigzag or armchair) directions of both A-A and A-B stacked bilayer graphene.

One of the technological drivers of GNR science is the desire to realise scalable sub-10nm electronics[48]. Here we demonstrate that as the 3nm length scale is approached, a transition to a new regime occurs, in which the physics of GNRs and other shapes is not simply a downscaling of larger structures, because the proximity of one EC to another causes the global reconstruction of whole structures and not simply their edges. This global reconstruction enables the creation of new molecular-structures, which cannot be synthesised by any other method.

The aim of this Letter is to establish the principle of this new methodology. We demonstrate that by sculpting selected shapes from *bilayer* graphene and allowing the shape to reconstruct globally, unique sp$^2$-bonded molecular structures can be formed and a variety of multiply-connected sp$^2$-bonded geometries can be created. This reconstruction is demonstrated using density functional theory and where appropriate, classical molecular dynamics (see supporting information). As a first example of such *sculpturenes*, we demonstrate that sculpted bilayer graphene can reconstruct to form new sp$^2$-bonded carbon cages, which are topologically distinct from fullerenes. Consider cutting a hexagonal shape in bilayer graphene (located on top of a graphene substrate), using for example STM lithography[29], to form the AB-stacked bilayer annulus, shown in figure 1, where atoms of the upper (lower) layer of the nanoribbon are shown in red (blue) and atoms belonging to the graphene substrate are shaded grey. Figures 1b and 1c show two views of the resulting toriodal sculpturene, which is formed after allowing this ribbon to relax and spontaneously reconstruct (for clarity, the substrate is not shown).

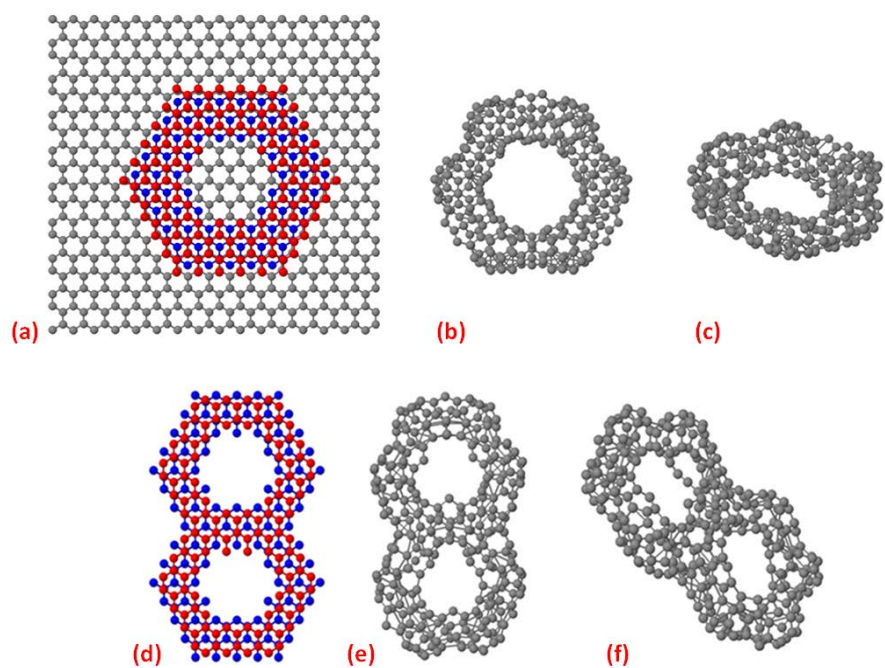

**Figure 1.** Figure 1a shows a hexagonal bilayer ribbon stacked on a graphene substrate, obtained by sculpting the top two layers of AB stacked graphene, where A refers to the top sheet (red) and B refers to the sheet immediately beneath the top sheet (blue). Figures 1b and 1c show different views of the toroidal sculpturene obtained by allowing the GNR of fig 1a to reconstruct (for clarity, the graphene substrate is not shown). The outer (inner) diameter of this toroidal scupturene is approximately 2.2nm (1.1nm). Figure 1d shows a double-hexagonal shape (283 carbon atoms) sculpted from AB stacked graphene. Figures 1e and 1f show different views of the resulting sculpturene with connection k=5, obtained by allowing the initial shape to reconstruct.

Unlike other pure forms of $sp^2$ carbon, such as fullerenes and closed carbon nanotubes, which are topologically equivalent to a sphere and characterised by an order of connection k=1, the above toriodal sculpturene possesses a connection of k=3. (The order of connection k is 1 plus the number of closed cuts that can be made on a given surface without breaking it apart into two pieces[39].) Since the connection k is a topological invariant, toroidal sculpturenes cannot be continuously deformed into fullerenes or nanotubes. Furthermore, since k is unchanged by the introduction of atomic-scale defects and impurities, this method of constructing new topologies is extremely robust.

In contrast with conventional fullerenes, the $sp^2$-bonded torus shown in figures 1b and 1c is expected to exhibit interesting orbital magnetic effects, including persistent currents[33-38], which are a direct consequence of its topology. Furthermore, the presence of a hydrophobic pocket in the centre of the torus provides a unique binding site for supra-molecular attachment to other molecules.

The toriodal sculpturene of figures 1b, c is the simplest example of a hierarchy of $sp^2$-bonded toroidal molecular structures with higher connections. As a second example, figure 1d shows a sculpturene formed from two hexagons, which reconstructs to form the "figure of eight" shown in figures 1e,f, which possesses a connection k=5. The toriodal sculpturene of figures 1b, c is of order 100 times smaller than ring structures reported elsewhere in the literature, which are either formed accidentally[40,41] or by etching CNTs to form loops with diameters of order 0.5 microns[42]. Toriodal molecular structures with k greater than 3, such as that shown in figures 1e and 1f are unprecedented and cannot be created by any other method.

This method of sculpting bilayers and allowing them to reconstruct is very versatile and can be used to create multiply-connected forms of $sp^2$-bonded carbon in a deterministic manner. Figures 2a and 2b show an example of a T-branched geometry and figures 2c and 2d an example of a cross geometry, obtained by allowing their corresponding sculpted bilayers to reconstruct. These hollow structures could form the basic building blocks of future nanofluidic devices.

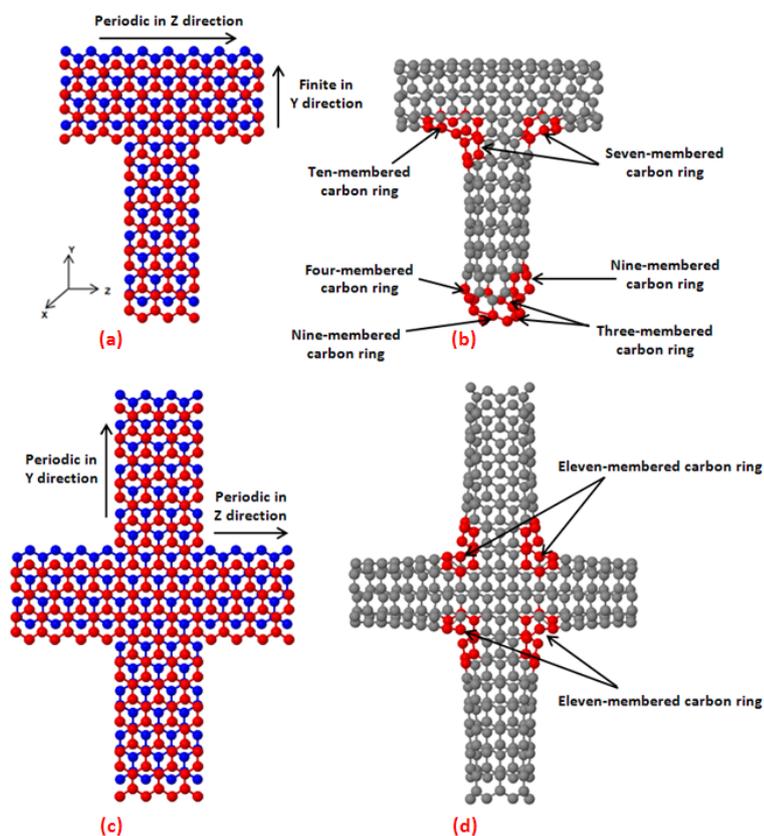

**Figure 2.** (a) A T-branched structure made by sculpting an AB-stacked bilayer. The structure is periodic in the horizontal direction. (b) The resulting T-branched CNT structure obtained by relaxing the bilayer ribbon of figure 2a. (c) A cross-like structure made by sculpting an AB-stacked bilayer. The structure is periodic in the horizontal and vertical direction. (d) A relaxed cross-like carbon nanotube structure obtained by relaxing the bilayer ribbon in figure 2c.

In the supporting information and reference [8], it is demonstrated that simple rectangular nanoribbons sculpted from bilayer graphene can also reconstruct to form CNTs with a predetermined chirality. At first sight, the spontaneous formation of CNTs from bilayer graphene nanoribbons is counterintuitive, because this is the reverse of a known process in which CNTs can unzip to form GNRs[18-24]. However the experiments demonstrating the latter were performed on relatively wide-diameter CNTs, whereas in the supporting information, we now show that the formation of CNTs from free-standing bilayer graphene nanoribbons is energetically-favoured only for nanoribbons of width less than 3nm, in agreement with ref [33]. If the nanoribbons are greater than this width, then they remain flat even though their edges can reconstruct to maximise $sp^2$ bonding. Similarly the global reconstruction leading to the above toroidal scultpturenes and multiply-connected structures occurs when the width of the BiGNR is of order 3nm or less and therefore the inner and outer ECs are in close proximity. Wider BiGNRs would possess ECs, but the cross-sections of their interiors would assume a 'dog-bone' shape, as shown in figure S2 of the supporting information.

Sculpturenes can also be formed from different materials. As an example, figure 3 shows a hetero-bilayer formed from a graphene sheet, AB stacked onto a boron nitride sheet. After sculpting the bilayer to form a nanoribbon and allowing the ribbon to relax, the sculpturene on the right is formed, which comprises hetero-nanotube with one half formed from carbon and the other from boron nitride. Hetero-toriodal sculpturenes can also be formed in this way.

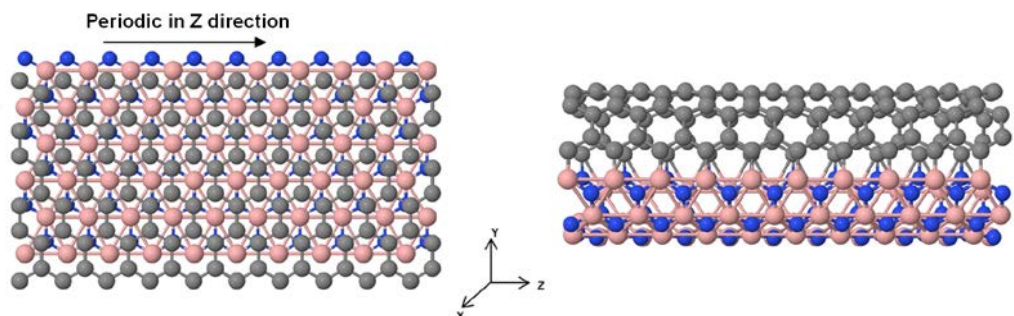

**Figure 3.** Left: A Super Cell of hetero-bilayer nanoribbons of boron nitride/graphene which contains 240 atoms. Right: A hetero-CNT obtained by relaxing the structure on the left.

Having demonstrated the formation of molecular structures and topologies from sculpted bilayer ribbons of figure 1-3, which are of approximately uniform width, we now turn to structures with variable widths, and show that these can yield CNTs with a chosen chirality, at a specific location, whose ends are automatically connected to graphene electrodes. As an example, we have relaxed a sculpted free-standing AB-stacked bilayer formed by periodically repeating the unit cell shown on the left in figure 4a in the horizontal and vertical directions, which consists of a narrow central bilayer ribbon in contact with two wider bilayer ribbons. We find that this unit cell spontaneously relaxes to the finite-length CNT shown on the right in figure 4b, which is connected on either side to sheets of bilayer graphene.

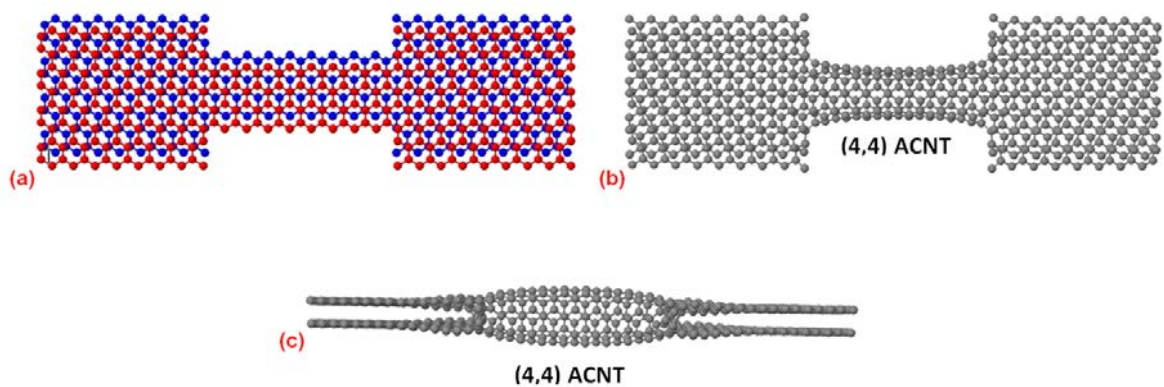

**Figure 4.** Fig 4a shows a finite AB-stacked BiGNR whose ends are connected to bilayer graphene electrodes; the unit cell contains 644 carbon atoms. Figs 4b and 4c show two views of the structure after relaxation, which comprised a CNT in the central region, attached to bilayer graphene 'electrodes'.

Clearly the chirality of the CNT in the central region of figure 4c is determined by the crystallographic direction of the lines cut into the initial bilayer. (See supporting information for further examples.) As expected, only the narrow inner part of the above structure (of width less than 3nm) relaxes to form a CNT, whereas the wider outer sections remain planar.

Since stacked ECs in 4-layer graphene have been observed[45] it is natural to ask if sculpted stacks of n monolayers, with n>2, leads to sculpturenes, which are stacked perpendicular to the plane of the monolayers. As an example, figure 5 shows the result of cutting parallel lines in AA stacked 4-layer graphene to yield the 4-layer nanoribbon (on the left) and then and allowing the structure to relax.

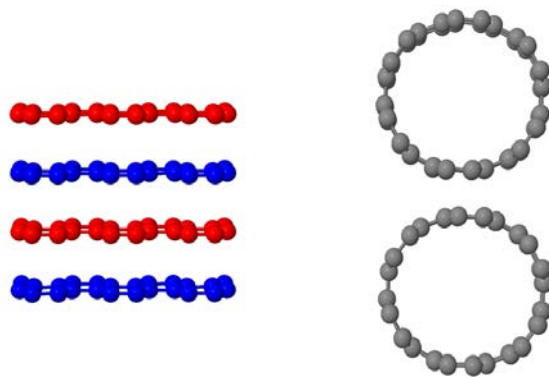

**Figure 5.** The left figure shows a supercell of AA-stacked 4-layer graphene, which contains 96 carbon atoms. The right figure shows that the structure relaxes to two aligned CNTs.

The resulting sculpturene (on the right) consists of two parallel-aligned CNTs, stacked in a direction which is perpendicular to the planes of the original monolayers. This example demonstrates that three-dimensional patterning of sculpturenes is possible.

Finally we show that hollow multiply-connected structures with potential to support nanofluidic transport can be formed by sculpting appropriate BiGNRs and allowing them to reconstruct. Figure 6 shows one example of a basic building block, in which the relaxed structure comprises a horizontal nanotube (on the left), which bifurcates into two parallel channels. These channels later rejoin and attach to seamlessly to a nanotube on the right.

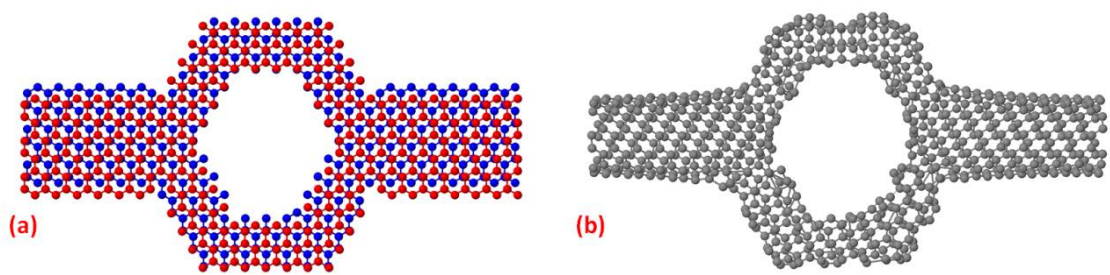

**Figure (6) Formation of a hollow torus connected to CNTs.** A supercell of AB-stacked BiGNR, which contains (676 carbon atoms), before (a) and after (b) reconstruction.

In summary, we have demonstrated that sculpted bilayer graphene and hetero-bilayers, can reconstruct spontaneously to form unique forms of $sp^2$-bonded molecular structures, which have dimensions comparable with those of conventional fullerenes and nanotubes. The toriodal sculpturene of figures 1b,c is of order 100 times smaller than ring structures reported elsewhere in the literature, which are either formed accidentally[40,41] or by etching CNTs to form loops with diameters of order 0.5 microns.[42] Furthermore, the sculpturene of figures 1e,f has an unprecedented connection k > 3. During recent years, there have been several studies of edge morphology in both AB and AA-stacked graphene layers and rapid progress is being made in achieving control at an atomistic level[44-46]. Even without such control, the sculpting of bilayers to create new topologies is robust, because topological invariants are independent of the detailed atomic structure. The presence of a hydrophobic pocket in the centre of the torus is also a robust topological feature, which has potential supra-molecular sensing applications, particularly when such a structure is embedded in nano-circuitry.

By sculpting bilayer graphene in this way, we have demonstrated that multi-terminal sculpturenes with non-trivial shapes can be formed deterministically, including T-junctions, crosses and toruses. The method can also be used to select the chirality of CNTs, place them at desired locations within a device and ohmically contact them to electrodes in a planar geometry. Furthermore the method can be used to produce hetero-nanotubes. As an example, we examined the spontaneous formation of nanotubes obtained by sculpting a hetero-bilayer from a monolayer of graphene on top of a single layer of boron nitride and find that the resulting nanotubes consist of a half cylinder of carbon joined to a half cylinder of boron nitride. Clearly the method can be generalised to create hetero-bilayers formed from other combinations of layered materials and hetero-sculpturenes with non-trivial topologies.

In the supporting information, it is noted that for free-standing GNRs, global reconstruction occurs only when ECs are separated by of order 3nm or less. For the future it will be of interest to develop methods for promoting the global reconstruction of wider GNRs. One such method is likely to involve chemically-coating the GNRs during or after sculpting, since this is known to promote the rolling of flat graphene into curved surfaces[8]. Even without such assistance, the size of sculpturenes is not restricted to the nanoscale and is limited only by the size of the initial bilayer and the ability to cut the bilayer. For example figure S7 of the supporting information shows a toroidal sculpturene, which is much larger than that of figures 1b and 1c formed by relaxing a hexagonal annular bilayer comprising 475 carbon atoms. The six corners of this larger toroidal sculpturene are fullerene-like chambers, whereas the straighter regions connecting the corners are CNTs. Networks of fullerene-like chambers connected by CNTs could form a basis for a nanofluidic devices in which molecules can flow between the chambers via the CNT interconnects of the kind shown in figures 2 and 6.

At present, STM lithography[11] can be used for the top-down fabrication of GNRs and other shapes with length scales less than 3nm, although this suffers from low throughput. Other methods are continuously evolving. For example, techniques for cutting of graphene using nanoparticles have been developed[12,13,49], although these currently create graphene shapes with dimensions greater than 10nm. Chemical unzipping of CNTs also produces GNRs wider than 10nm[19,50,51]. On the other hand a chemical method for bottom up fabrication of GNRs with widths less than 3nm has been demonstrated[25], although this has not yet generated BiGNRs.

Since GNRs commonly contain defects such as vacancies, 5-7 pairs of rings, loops and interstitials, defect edge engineering is needed to control the reactivity and properties of graphene edges[46]. The SI shows that CNT sculpturenes can be either defect free or contain ordered arrays of non-hexagonal ring structures. Rows of 5-8-5 rings have been observed within a graphene sheet using STM[52] and can behave as quantum wires. Arrays of non-hexagonal rings in CNT sculpturenes are similarly expected to possess interesting electronic properties.

**Acknowledgements**

This work is supported by the European Union Marie-Curie Networks FUNMOLS and NANOCTM, the UK EPSRC and the Iraqi Ministry of Higher Education and Scientific Research.